\documentclass[aps,prl,reprint,showpacs]{revtex4-1}
\usepackage{graphicx} 
\usepackage{amsmath} 
\usepackage{bm} 
\usepackage{natbib}
\usepackage{natmove}
\usepackage[colorlinks=true, linkcolor=blue, citecolor=blue, urlcolor=blue, linktoc=page, bookmarks=false, pdfstartview={FitH}, pdfborder={0 0 0.0 [3 3]}]{hyperref} 
\usepackage{color} 
\usepackage{dcolumn} 
\newcolumntype{.}{D{.}{.}{2.1}}
\newcolumntype{-}{D{.}{.}{4.0}}

\begin{document}

\title{Tuning ferromagnetism at interfaces between insulating perovskite oxides}
\author{Nirmal Ganguli}
\email[Email: ]{N.Ganguli@utwente.nl}

\author{Paul J.\ Kelly}
\email[Email: ]{P.J.Kelly@utwente.nl}
\affiliation{Faculty of Science and Technology and MESA$^+$ Institute for Nanotechnology, University of Twente, P.O.\ Box 217, 7500 AE Enschede, The Netherlands}


\begin{abstract}
We use density functional theory calculations to show that the LaAlO$_3|$SrTiO$_3$ interface between insulating perovskite oxides is borderline in satisfying the Stoner criterion for itinerant ferromagnetism and explore other oxide combinations with a view to satisfying it more amply. The larger lattice parameter of an LaScO$_3|$BaTiO$_3$ interface is found to be less favorable than the greater interface distortion of LaAlO$_3|$CaTiO$_3$. Compared to LaAlO$_3|$SrTiO$_3$, the latter is predicted to exhibit robust magnetism with a larger saturation moment and a higher Curie temperature. Our results provide support for a ``two phase'' picture of coexistent superconductivity and ferromagnetism.
\end{abstract}
\pacs{75.70.Cn, 73.20.-r, 71.28.+d}
%
%
%
\maketitle

{\color{red}\it Introduction.}---LaAlO$_3|$SrTiO$_3$ (LAO$|$STO) heterostructures have received a great deal of attention over the past decade following the observation of a high mobility two dimensional electron gas (2DEG) at the interface between the two band insulators \cite{Ohtomo:nat04}. Even more intriguing is the finding that superconductivity and ferromagnetism coexist (up to 100-120~mK) \cite{Dikin:prl11,Bert:natp11,Li:natp11} where neither material on its own exhibits ferromagnetism; doped bulk STO is known to be superconducting \cite{Schooley:prl64}. Though magnetic ordering \cite{Brinkman:natm07} is now established, the size of the magnetic moments and the ordering temperature are very sensitive to details of how the interfaces are prepared and how the magnetism is measured \cite{Ariando:natc11,Dikin:prl11,Li:natp11,Bert:natp11,Fitzsimmons:prl11,Kalisky:natc12,Salman:prl12,Lee:natm13}. Magnetic torque magnetometry measurements found a saturation moment of $\sim$0.3~$\mu_B$ per interface unit cell with magnetization persisting above 200~K \cite{Li:natp11} confirming signatures of room temperature ferromagnetism reported by \citet{Ariando:natc11}. Scanning superconducting quantum interference device measurements revealed submicrometer patches of ferromagnetism on a uniform paramagnetic background \cite{Bert:natp11}. However, other experiments fail to observe significant interface magnetization \cite{Fitzsimmons:prl11,Salman:prl12} suggesting that its occurrence depends on the experimental conditions during sample preparation and measurement.

\begin{figure}[b]
\includegraphics[scale = 0.34]{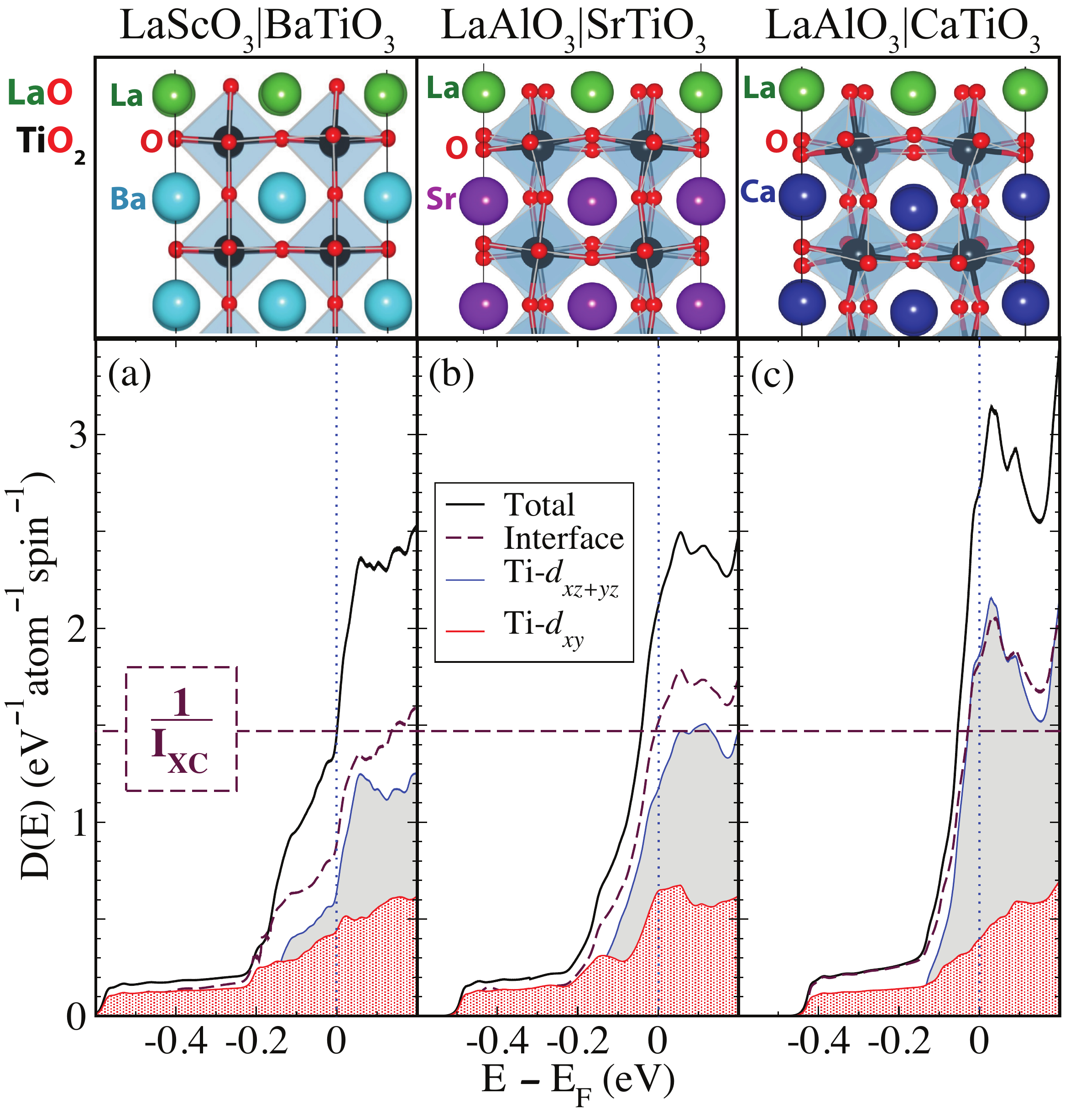}
\caption{\label{fig:strDoS}(Color) Minimum energy interface structures (top) and corresponding non-magnetic DoS $D(E)$ for (a) LaScO$_3|$BaTiO$_3$, (b) LaAlO$_3|$SrTiO$_3$, (c) LaAlO$_3|$CaTiO$_3$. The in-plane lattice constant decreases from left to right. As well as the total DoS, we show $D(E)$ resolved into $d_{xy}$ and $d_{xz,yz}$ components and also projected onto the interface TiO$_2$ layer. The horizontal dashed line is $I^{-1}_\text{xc}$ for Ti.
}
\end{figure}
 
The sensitivity of the interface magnetism observed in experiment is reflected in density functional theory (DFT) calculations where magnetic ordering depends on the choice of exchange-correlation potential and details of how the LAO$|$STO interface structure is modelled \cite{Pentcheva:prb06,Janicka:jap08,Zhong:epl08,Li:prb13}. This sensitivity suggests that the interface Ti-$d$ bands may be very close to a magnetic instability resulting from competition between kinetic and exchange energy. This competition is conventionally formulated as the Stoner criterion, $D(E_F) I_\text{xc} \ge 1$, in terms of the non-magnetic density of states (DoS) at the Fermi energy $D(E_F)$, and the Stoner parameter $I_\text{xc}$ that has been evaluated in the local spin density approximation (LSDA) to DFT and describes correctly the occurrence of itinerant ferromagnetism for metals \cite{Gunnarsson:jpf76,Poulsen:jpf76,Janak:prb77}. 

The proximity to a magnetic instability is confirmed in Fig.~\ref{fig:strDoS} where  $D(E)$ and $I^{-1}_\text{xc}$ are plotted for a number of $n$-type interfaces formed from bulk materials with lattice parameters larger and smaller than that of SrTiO$_3$ and with TiO$_6$ rotations that are smaller and larger than predicted for LAO$|$STO \cite{Zhong:epl08}. The figure shows that when the interface Ti $d_{t_{2g}}$ band contains half an electron, the Stoner criterion is not satisfied for LaScO$_3|$BaTiO$_3$ but is satisfied for LaAlO$_3|$CaTiO$_3$; increasing the lattice constant is much less effective in increasing $D(E_F)$ than rotating the TiO$_6$ octahedra. The figure also makes clear how essential adequate band-filling is for realizing magnetism.  

Motivated by the interpretation of the theoretical \cite{Pentcheva:prb06,Janicka:jap08,Zhong:epl08,Li:prb13} and experimental \cite{Ariando:natc11,Dikin:prl11,Li:natp11,Bert:natp11,Fitzsimmons:prl11,Kalisky:natc12,Salman:prl12,Lee:natm13} results afforded by Fig.~\ref{fig:strDoS}(b), we investigate how to make the interface magnetism more robust. Because the Stoner parameter is essentially a fixed, atomic property of Ti, we focus on how to increase $D(E_F)$: by increasing $D(E)$ as a whole, by changing the band filling to shift $E_F$ to a position of higher state density, or by a combination of both.

The DoS is inversely proportional to the bandwidth. It can be increased by rotating the TiO$_6$ octahedra \cite{Zhong:epl08} or by increasing the in-plane lattice constant by replacing STO with the larger BaTiO$_3$ (BTO) that has a lattice constant $a = 4.00$~\AA\ in its cubic form. To avoid problems related to strain, LAO should be replaced with an A$^{3+}$B$^{3+}$O$_3$ oxide with a larger, matching lattice constant. We consider the recently synthesized large bandgap scandates, YScO$_3$ (YSO) and LaScO$_3$ (LSO) \cite{Balamurugan:znb10} that retain the $+/-$ charged layers alternating along the $\langle 001 \rangle$ direction that dopes the interface while offering a good lattice match to BTO; the pseudocubic lattice constants of orthorhombic YSO and LSO are 3.94~\AA\ and 4.05~\AA, respectively (see Table~\ref{tab:LatConst}).
The second possibility we consider is to replace STO with the smaller CaTiO$_3$ (CTO) \cite{Howard:acsb98, Sasaki:acsc87} that may favour larger tilt angles at the interface; CTO is an orthorhombic compound where the TiO$_6$ octahedra are intrinsically tilted. Its pseudocubic lattice constant of 3.80~\AA\ matches that of LAO, 3.79~\AA, almost perfectly. 

\begin{table}[b]
\caption{\label{tab:LatConst} Orthorhombic lattice parameters $a$, $b$ and $c$ and pseudocubic lattice parameter $\hat{a}=\sqrt[3]{abc/4}$ of the six perovskite structure materials considered in this paper in \AA. For the cubic materials LAO, STO and BTO, $\hat{a} \equiv a$.}
\begin{ruledtabular}
\begin{tabular}{rllll|llllll}
           & \multicolumn{4}{c|}{Experimental}   
                                   &  \multicolumn{4}{c}{Calculated} \\
                     \cline{2-5}                   \cline{6-9}
           & \multicolumn{1}{c}{$a$}  
                  & \multicolumn{1}{c}{$b$} 
                           & \multicolumn{1}{c}{$c$} 
                                   & \multicolumn{1}{c|}{$\hat{a}$}                           
                                           & \multicolumn{1}{c}{$a$}  
                                                   & \multicolumn{1}{c}{$b$} 
                                                          & \multicolumn{1}{c}{$c$} 
                                                                 & \multicolumn{1}{c}{$\hat{a}$} \\
\hline
LaAlO$_3$  & 3.79  & 3.79  & 3.79  & 3.79  &  3.74 & 3.74 & 3.74 & 3.74 \\
CaTiO$_3$  & 5.380 & 5.442 & 7.640 & 3.80  &  5.30 & 5.43 & 7.55 & 3.79 \\
SrTiO$_3$  & 3.905 & 3.905 & 3.905 & 3.905 &  3.89 & 3.89 & 3.89 & 3.89 \\
 YScO$_3$  & 5.707 & 7.893 & 5.424 & 3.94  &  5.70 & 7.87 & 5.39 & 3.92 \\
BaTiO$_3$  & 4.00  & 4.00  & 4.00  & 4.00  &  3.97 & 3.97 & 3.97 & 3.97 \\ 
LaScO$_3$  & 5.797 & 8.103 & 5.683 & 4.05  &  5.78 & 8.05 & 5.66 & 4.04 \\
\end{tabular}
\end{ruledtabular}
\end{table}

{\color{red}\it Method.}---YSO$|$BTO, LSO$|$BTO, LAO$|$STO and LAO$|$CTO interfaces are constructed by total energy minimization within DFT. All structural optimization and electronic structure calculations were carried out with a projector augmented wave (PAW) basis \cite{paw} as implemented in \textsc{vasp} \cite{Kresse:prb93, *Kresse:prb96}, the local spin density approximation (LSDA) as parameterized by Perdew and Zunger \cite{Perdew:prb81} combined with an on-site Hubbard $U$ (LSDA+$U$) to correct the underestimation of $d$ electron localization in the LSDA \cite{Dudarev:prb98}. Unless otherwise stated, a moderate value of $U-J = 3$~eV is used for the Ti-$3d$ states that gives a good description of the structural and magnetic properties of bulk $3d^1$ oxides \cite{Ganguli:d1Oxide}. Spectral properties are calculated using a value of $U-J = 10$~eV on the La $4f$ states which would otherwise lie too low in energy compared to the Ti $3d$ states \cite{Okatov:epl05}. 
The atomic positions are relaxed to minimize the Hellman-Feynman forces on each atom with a tolerance value of 0.01~eV/\AA. 

Our starting point is a bulk titanate substrate that is assumed to be so thick that it determines the in-plane lattice constant. To model, for example, an LAO$|$STO interface, we use a periodically repeated ($m, n$) supercell containing $m$ unit cells of LAO and $n$ unit cells of STO perpendicular to the interface. The results reported in this paper were obtained with a ($\frac{5}{2},\frac{15}{2}$) supercell containing two $n$-type interfaces and an in-plane $p(2\times2)$ unit cell to enable full rotational freedom of the TiO$_6$ interface octahedra. Apart from constraining the in-plane lattice constants to the titanate bulk value, all structural parameters of these 200 atom supercells including the out-of-plane lattice parameter were optimized, representing a substantial improvement on previous calculations. 


{\color{red}\it Results.}---The structures of the calculated lowest energy interfaces are shown in the upper panels of Figure~\ref{fig:strDoS}. From left to right, it is apparent that the rotation of the interface TiO$_6$ octahedra increases dramatically on going from a BTO to an STO to a CTO substrate; from the structure with the {\it largest} in-plane lattice constant to that with the {\it smallest} \footnote{We find an $a^+b^-c^-$ tilt system for the TiO$_6$ octahedra near the LAO$|$CTO interface as opposed to an $a^+b^-b^-$ tilt system for bulk CaTiO$_3$ \cite{Woodward:acsb97b}.}. For interfaces that are allowed to relax but not to rotate, the DoS decreases in this sequence. However, the lower panels of Fig.~\ref{fig:strDoS} show that the opposite is true in the unconstrained case. There, $D(E_F)$ increases from 1.5 to 2.1 to 2.7 states per eV~atom~spin as the octahedral rotation narrows the interface Ti $t_{2g}$ band \cite{Zhong:epl08}. For magnetism to occur, it is the DoS projected onto the interface Ti atoms that is relevant; it increases from 0.9 to 1.5 to 1.8. For comparison, the horizontal line in Fig.~\ref{fig:strDoS} represents $I^{-1}_{\rm xc}$ where for Ti, $I_{\rm xc} \sim 0.68$~eV \cite{Janak:prb77,Andersen:85}. It is clear that the Stoner criterion is not satisfied for LSO$|$BTO, is borderline for LAO$|$STO, and is amply satisfied by LAO$|$CTO. 

Also apparent from the figure is the different role played by Ti $d_{xy}$ electrons that are localized at the interface and highly dispersive in the interface plane, and the $d_{xz}$ and $d_{yz}$ electrons that have a very anisotropic in-plane dispersion but extend further into the titanate substrate \cite{Popovic:prl08,Delugas:prl11}. The bottom of the interface band is formed by the $d_{xy}$ electrons but the corresponding DoS is much too low for these states to order magnetically. The $d_{xz,yz}$ states have a much higher DoS which increases greatly when the TiO$_6$ octahedra rotate, unlike the $d_{xy}$ electrons. The sharp increase in the $d_{xz,yz}$ DoS at $E_F$ highlights the importance of having a sufficient number of electrons in the interface bands and the possibility of tuning the magnetism by field doping with relatively small numbers of electrons as well as suggesting other strategies for inducing ferromagnetism in LAO$|$STO by enhancing $D(E_F)$, e.g., with strain.

\begin{figure}
	\includegraphics[scale = 0.32]{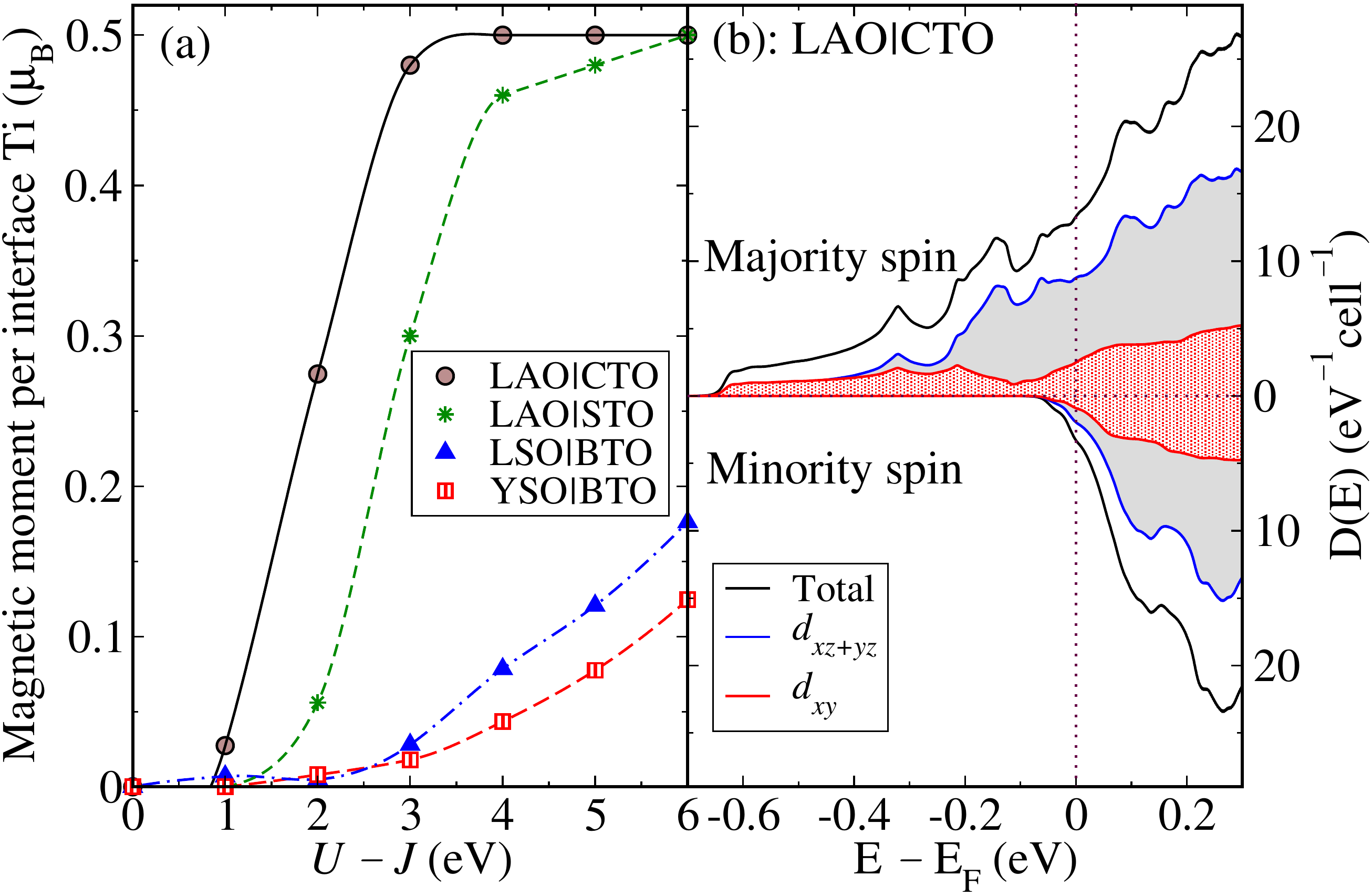}
	\caption{\label{fig:magUdos}(Color online) (a) Variation of the magnetic moment with value of $U-J$. (b) Spin polarized density of states for LAO$|$CTO for $U-J=3$~eV.}
\end{figure}

{\color{red}\it Magnetism.}---The Stoner criterion signals a magnetic instability but determining the equilibrium magnetization requires iteration to self consistency \footnote{For inhomogeneous systems, the Stoner criterion is at best indicative because of the arbitrariness in defining density of state projections and because the exchange polarization in complex materials more often than not does not have rigid-band character.}. The influence of $U$ on the magnetic moments for all four interfaces is shown in Fig.~\ref{fig:magUdos}(a). The YSO$|$BTO (LSO$|$BTO) interfaces only form appreciable magnetic moments for large values of $U$, reaching values of only 0.12 (0.17)~$\mu_B$ per interface Ti atom for $U-J = 6$~eV. 

The LAO$|$STO magnetic moment increases rapidly for $U-J>2$~eV and attains a value of 0.3~$\mu_B$ per interface Ti, the largest value reported \cite{Li:natp11}, for $U-J=3$~eV; this corresponds to a value of $U$ in the range reported to give good agreement with experiment for bulk LaTiO$_3$ \cite{Okatov:epl05,MizokawaPRB95}. For this value of $U-J$, the LAO$|$CTO magnetic moment has almost reached its saturation value of 0.5~$\mu_B$ after a rapid increase for $U-J > 1$~eV. 
The corresponding density of states in Fig.~\ref{fig:magUdos}(b) confirms the essentially complete spin polarization of the conduction band electrons, a feature that could be important if spin dependent transport can be demonstrated. The self-consistent results confirm the Stoner-criterion picture that YSO$|$BTO and LSO$|$BTO interfaces are unlikely to host ferromagnetism, while LAO$|$CTO is expected to be a more promising candidate than LAO$|$STO. 

\begin{figure}
	\includegraphics[scale = 0.35]{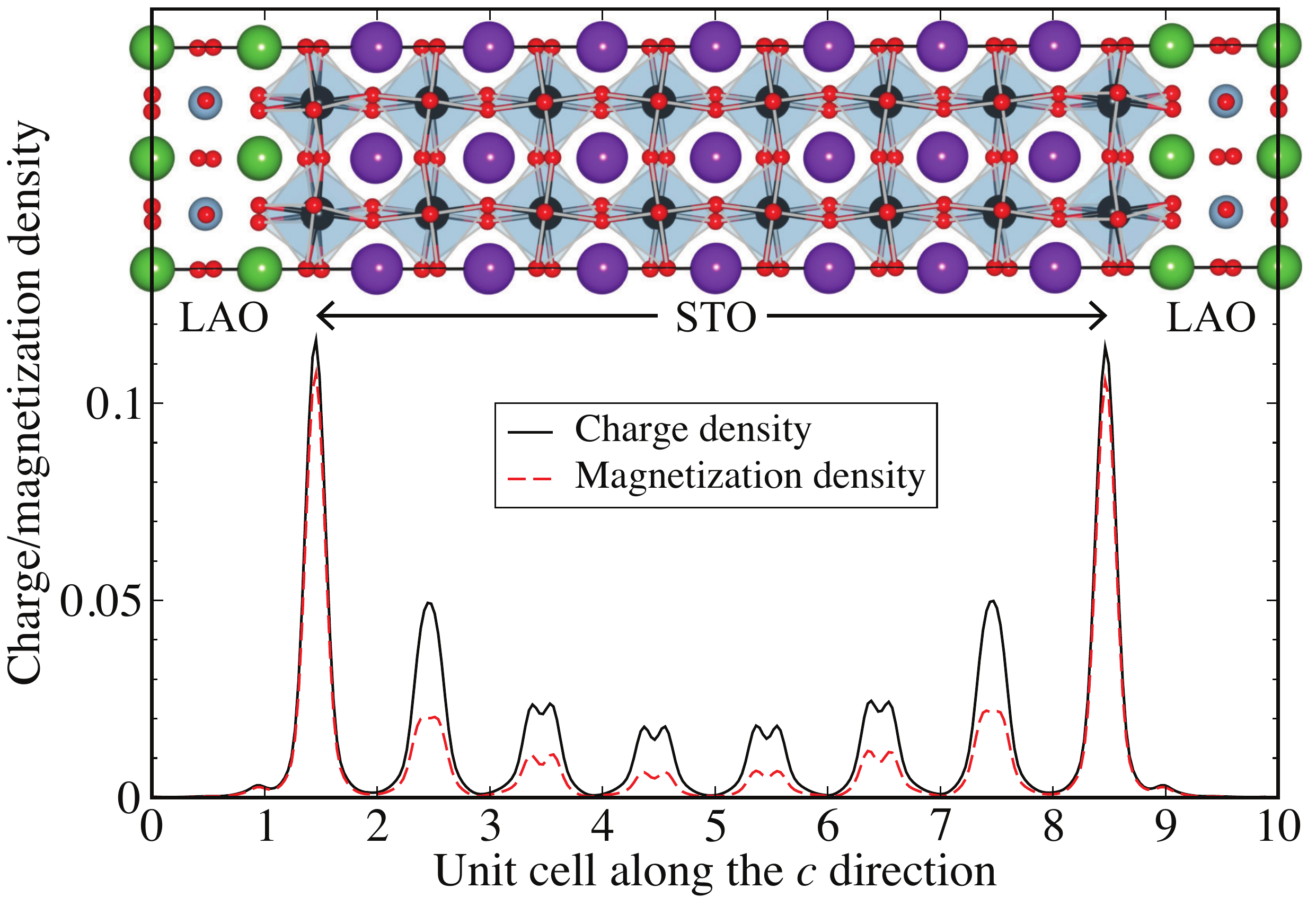}
	\caption{\label{fig:Ch&Sp}(Color online) Plane-averaged charge and magnetization densities for LAO$|$STO calculated with $U-J=3$~eV.}
\end{figure}

Comparison of the DoS in Figs.~\ref{fig:magUdos}(b) and \ref{fig:strDoS}(c) underlines the non-rigid nature of the spin-polarization. Where the $d_{xz,yz}$ states are largely responsible for the initial magnetic instability, it is the $d_{xy}$ states that subsequently profit most from it \cite{Lee:natm13}. This is because the latter are highly localized in the interface layer and polarize almost completely; their large exchange splitting of some 0.6~eV localizes the occupied states in the interface layer even more. For LAO$|$CTO, the complete spin-polarization means that the interface charge and spin densities coincide. 
When, as happens for LAO$|$STO, the initial nonmagnetic DoS is lower, Fig.~\ref{fig:strDoS}(c), the degree of spin polarization is less complete. In this case the interesting situation shown in Fig.~\ref{fig:Ch&Sp} arises where the charge density near the interface has largely $d_{xy}$ character and is almost completely spin polarized, while there is a substantial charge density extending out into the STO with $d_{xz,yz}$ character that is only weakly polarized. We identify this with the ``two independent carrier gases'' proposed by Dikin et al.  to explain the observed coexistence of superconductivity and ferromagnetism \cite{Dikin:prl11} and ``two phase'' scenario of Ariando et al. \cite{Ariando:natc11}.
\begin{table}[b]
	\caption{\label{tab:magOrder}Energies for nonmagnetic (NM), ferromagnetic (FM), striped antiferromagnetic (St-AFM), and checkerboard antiferromagnetic (CB-AFM) ordering for LAO$|$STO and LAO$|$CTO interfaces calculated with $U-J=3$~eV. Energies are relative to the lowest energy FM state and are in meV per interface Ti ion.}
	\begin{ruledtabular}
		\begin{tabular}{cdddd}
			& \multicolumn{1}{c}{NM} & \multicolumn{1}{c}{FM} & \multicolumn{1}{c}{St-AFM} & \multicolumn{1}{c}{CB-AFM} \\
			\hline
LAO$|$STO   &  6.9  & 0.0  & 2.7  & 1.3 \\		
LAO$|$CTO	& 11.3  & 0.0  & 9.3  & 5.2 \\
		\end{tabular}
	\end{ruledtabular}
\end{table}

{\color{red}\it Magnetic ordering.}---An important measure of the strength of magnetism is its ordering temperature. For ferromagnetic (FM) ordering, this is the Curie temperature $T_C$. In a Heisenberg model, $T_C$ is determined by the exchange coupling between atomic moments on different sites. For itinerant magnetism with only 0.5 electron per Ti ion, this picture is not applicable. Nevertheless, we can compare the stability of FM ordering to alternative types of antiferromagnetic (AFM) ordering by comparing the corresponding total energies. This is done in Table~\ref{tab:magOrder} for nonmagnetic (NM), FM, striped (St) and checkerboard (CB) AFM ordering of LAO$|$STO and LAO$|$CTO interfaces. The Table shows that the FM and NM states have, respectively, the lowest and highest energies and the AFM states are intermediate with CB ordering lower that St. If we assume that the persistence of FM up to room temperature for LAO$|$STO interfaces has been established \cite{Ariando:natc11}, then our total energies suggest that the FM coupling is much stronger for LAO$|$CTO and we expect magnetic ordering to persist to even higher temperatures.

{\color{red}\it LAO thickness dependence.}---By considering a symmetric multilayer geometry \cite{Pentcheva:prb06,Janicka:jap08,Zhong:epl08,Li:prb13}, an interface charge of 0.5 electron per interface Ti, the amount needed to resolve the ``polar catastrophe'' \cite{Nakagawa:natm06}, is automatically realized. For an overlayer of LAO of variable thickness grown on a semiinfinite XTO (X = Ca, Sr, Ba) substrate, this degree of charge transfer can only be achieved asymptotically for an infinitely thick overlayer. The amount of interface charge deduced from transport measurements is far less than 0.5 electron \cite{McCollam:aplm14} and this is one of the outstanding puzzles presented by these interfaces. 

\begin{figure}
	\includegraphics[scale = 0.35]{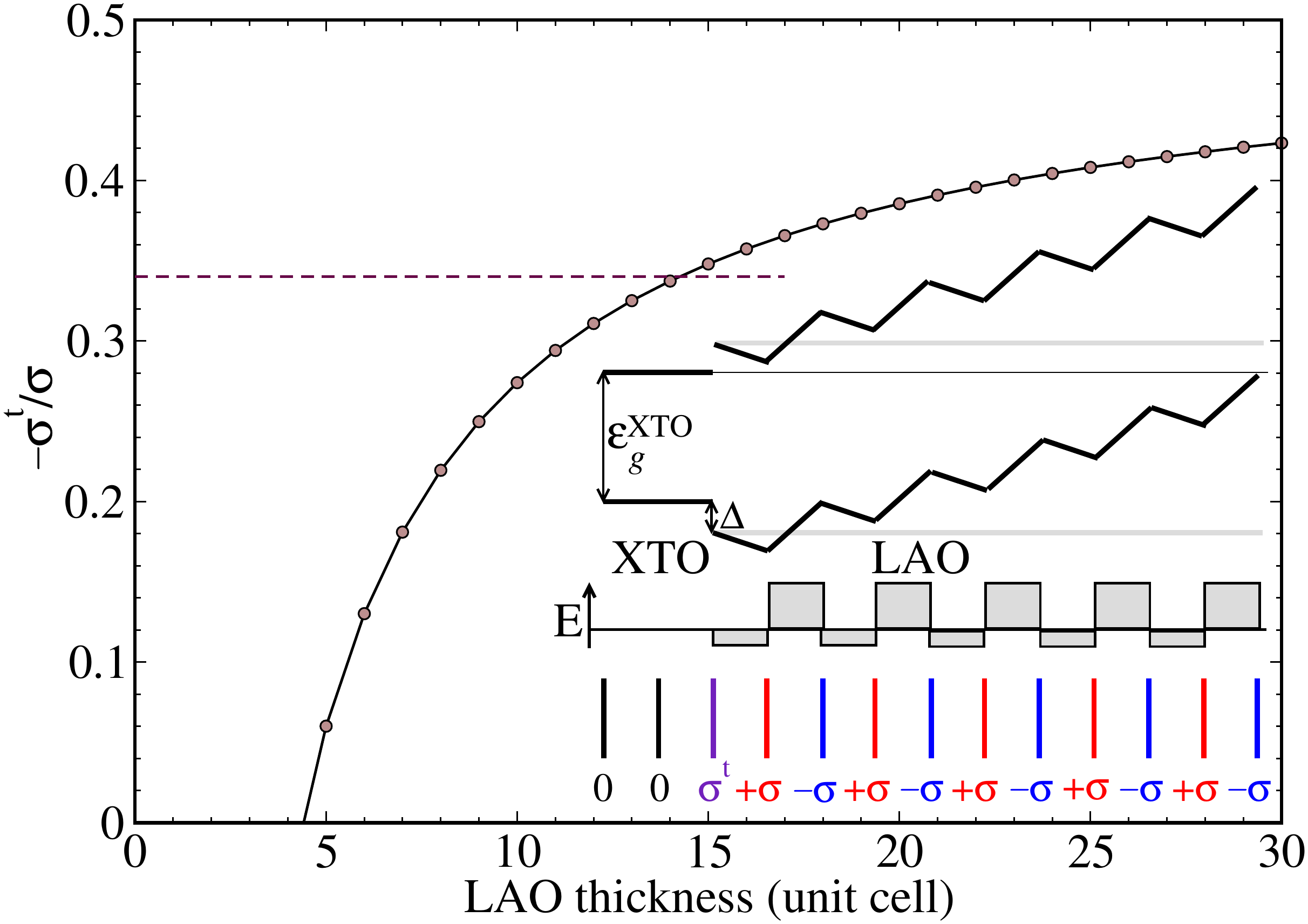}
	\caption{\label{fig:ChargeTransfer}(Color online) Charge transferred to the interface TiO$_2$ plane as a function of the number of LAO unit cells on bulk CTO calculated using Eq.~(\ref{eq:ChTr}). The horizontal dashed line indicates the amount of interface charge necessary to satisfy the Stoner criterion for LAO$|$CTO. Inset: illustration of the polar catastrophe.}
\end{figure}

$\sigma$ is the charge density of an LaO$^+$ plane and $\epsilon^\text{LAO}$ the permittivity of LAO.
A constant electric field $\sigma /\epsilon^\text{LAO}$ between LaO$^+$ and AlO$_2^-$ planes results in a potential build up across LAO that is proportional to the LAO thickness $nd$ measured in terms of $n$ LAO unit cells of height $d$ along the $\langle 001 \rangle$ direction; see the inset to Fig.~\ref{fig:ChargeTransfer}.  
According to the polar catastrophe scenario \cite{Nakagawa:natm06}, as more layers of LAO are added, its valence band rises until it coincides with the lowest unoccupied states in the XTO conduction band that are $\varepsilon_g^\text{XTO} + \Delta$ higher in energy where $\varepsilon_g^\text{XTO} $ is the XTO band gap and $\Delta$ the valence band offset with LAO. Charge is then transferred from the surface to the interface leading to an interface charge density $\sigma^t$ that reduces the field across the LAO whose thickness must be increased to transfer more charge until 0.5 electron has been transferred to each interface Ti ion. When this has happened, $\sigma^t = -\frac{1}{2}\sigma$ and there is no potential buildup. $\sigma^t$ can be expressed  as
\begin{equation}
 \frac{\sigma^t}{\sigma} = 
 - \frac{1}{2 + \frac{2 \epsilon^\text{LAO}}{nd}K^\text{IF}} 
 + \frac{\varepsilon_g^{\rm XTO} + \Delta}{\sigma\left( K^\text{IF} + 
                               \frac{nd}{\epsilon^\text{LAO}} \right)} 
\label{eq:ChTr}
\end{equation}
where $K^\text{IF} \equiv d^\text{IF}/\epsilon^\text{IF}$ depends on the effective position $d^\text{IF}$ of the interface charges and an effective interface dielectric screening $\epsilon^\text{IF}$. 
The right hand side of Eq.~(\ref{eq:ChTr}) approaches $-\frac{1}{2}$ with increasing LAO thickness $nd$ as $1/n$. 

Fig.~\ref{fig:ChargeTransfer} shows the variation of $\sigma^t / \sigma$ with $n$ for LAO$|$CTO assuming $\epsilon^\text{LAO}/\epsilon_0 = 24$ \cite{Konaka:js91} and $\epsilon_0$ is the permittivity of free space, $\varepsilon_g^{\rm CTO} = 3.57$~eV \cite{Ueda:jpcm98}, and $ \Delta = 0.76$~eV, the value we extract from our calculations. $K^\text{IF}$ is set to a reasonable value of 0.5~F$^{-1}$ \footnote{We found no significant change in $\sigma^\text{t} / \sigma$ as a function of LAO thickness for $\left(\hat{a}^\text{CTO}/{2 \epsilon^\text{LAO}} = 0.89 \right) \geq K^\text{IF} \geq \left(\hat{a}^\text{CTO}/{2 \epsilon^\text{CTO}} = 0.11 \right)$~F$^{-1}$. $\epsilon^\text{CTO}/\epsilon_0 = 190$ at room temperature \cite{Linz:jcp58}.}. The figure shows that there is no charge transfer for LAO less than 5 unit cells thick. The horizontal line indicates that for $\sigma^t / \sigma < 0.34$, the Stoner criterion is not satisfied, see Fig.~\ref{fig:strDoS}(c), suggesting that a minimum of 14 unit cells of LAO must be grown on CTO to satisfy the Stoner criterion, many more than are needed to trigger conduction. This model explains the observation of a critical thickness of LAO for the onset of ferromagnetism at the LAO$|$STO interface \cite{Kalisky:natc12}, but not why the critical thickness should be the same for conduction and magnetism. It also suggests the possibility of changing the electron density at the LAO$|$CTO interface using top/back gate or polar adsorbates (similar to LAO$|$STO interfaces \cite{Thiel:sc06, Xie:natc11, Bi:arXiv13}) and thereby tuning the magnetic properties even below 14 unit cell thickness of LAO, but not why this has not been observed \cite{Kalisky:natc12}. We hope that the appeal of magneto-electronic devices based upon gated LAO$|$CTO heterostructures, where a gate voltage can be used to switch the magnetization at the LAO$|$CTO interface on or off, will stimulate further studies of this novel system.

{\color{red}\it Conclusion.}---We have used {\em ab~initio} calculations to explore how the ferromagnetism observed at LAO$|$STO interfaces might be made more robust by increasing the lattice constant or increasing the tilting of TiO$_6$ octahedra to narrow the Ti $d$ bands. Replacing STO with BTO and LAO with YSO or LSO fails to narrow the bands sufficiently and is less favourable than LAO$|$STO. Replacing STO with CTO leads to better lattice matching, greater octahedron tilting and substantial narrowing of the Ti-$d_{xz}$ and $d_{yz}$ bands making LAO$|$CTO a promising system to explore for interface magnetism. Our calculations indicate that the exchange coupling between Ti atoms is larger with CTO than with STO so substantially higher Curie temperatures are expected. The polar catastrophe model suggests that LAO should be at least 14 unit cells thick in order to realize ferromagnetism at the LAO$|$CTO interface, while only 5 unit cells is sufficient for the onset of conduction. Alternatively, the electrons could be supplied by gating. Our results demonstrate that the picture proposed to explain the high mobility 2DEG \cite{Popovic:prl08} is compatible with itinerant ferromagnetism \cite{Pentcheva:prb06,Janicka:jap08,Zhong:epl08,Li:prb13} and provides support for the ``two phase'' interpretation of the coexistence of superconductivity and magnetism \cite{Ariando:natc11,Dikin:prl11}.

{\color{red}\it Acknowledgment.}---This work was financially supported by the ``Nederlandse Organisatie voor Wetenschappelijk Onderzoek'' (NWO) through the research programme of ``Stichting voor Fundamenteel Onderzoek der Materie'' (FOM) and the supercomputer facilities of NWO ``Exacte Wetenschappen'' (Physical Sciences).

\begin{thebibliography}{44}%
\makeatletter
\providecommand \@ifxundefined [1]{%
 \@ifx{#1\undefined}
}%
\providecommand \@ifnum [1]{%
 \ifnum #1\expandafter \@firstoftwo
 \else \expandafter \@secondoftwo
 \fi
}%
\providecommand \@ifx [1]{%
 \ifx #1\expandafter \@firstoftwo
 \else \expandafter \@secondoftwo
 \fi
}%
\providecommand \natexlab [1]{#1}%
\providecommand \enquote  [1]{``#1''}%
\providecommand \bibnamefont  [1]{#1}%
\providecommand \bibfnamefont [1]{#1}%
\providecommand \citenamefont [1]{#1}%
\providecommand \href@noop [0]{\@secondoftwo}%
\providecommand \href [0]{\begingroup \@sanitize@url \@href}%
\providecommand \@href[1]{\@@startlink{#1}\@@href}%
\providecommand \@@href[1]{\endgroup#1\@@endlink}%
\providecommand \@sanitize@url [0]{\catcode `\\12\catcode `\$12\catcode
  `\&12\catcode `\#12\catcode `\^12\catcode `\_12\catcode `\%12\relax}%
\providecommand \@@startlink[1]{}%
\providecommand \@@endlink[0]{}%
\providecommand \url  [0]{\begingroup\@sanitize@url \@url }%
\providecommand \@url [1]{\endgroup\@href {#1}{\urlprefix }}%
\providecommand \urlprefix  [0]{URL }%
\providecommand \Eprint [0]{\href }%
\providecommand \doibase [0]{http://dx.doi.org/}%
\providecommand \selectlanguage [0]{\@gobble}%
\providecommand \bibinfo  [0]{\@secondoftwo}%
\providecommand \bibfield  [0]{\@secondoftwo}%
\providecommand \translation [1]{[#1]}%
\providecommand \BibitemOpen [0]{}%
\providecommand \bibitemStop [0]{}%
\providecommand \bibitemNoStop [0]{.\EOS\space}%
\providecommand \EOS [0]{\spacefactor3000\relax}%
\providecommand \BibitemShut  [1]{\csname bibitem#1\endcsname}%
\let\auto@bib@innerbib\@empty
\bibitem [{\citenamefont {Ohtomo}\ and\ \citenamefont
  {Hwang}(2004)}]{Ohtomo:nat04}%
  \BibitemOpen
  \bibfield  {author} {\bibinfo {author} {\bibfnamefont {A.}~\bibnamefont
  {Ohtomo}}\ and\ \bibinfo {author} {\bibfnamefont {H.~Y.}\ \bibnamefont
  {Hwang}},\ }\href {\doibase 10.1038/nature02308} {\bibfield  {journal}
  {\bibinfo  {journal} {Nature}\ }\textbf {\bibinfo {volume} {427}},\ \bibinfo
  {pages} {423} (\bibinfo {year} {2004})}\BibitemShut {NoStop}%
\bibitem [{\citenamefont {Dikin}\ \emph {et~al.}(2011)\citenamefont {Dikin},
  \citenamefont {Mehta}, \citenamefont {Bark}, \citenamefont {Folkman},
  \citenamefont {Eom},\ and\ \citenamefont {Chandrasekhar}}]{Dikin:prl11}%
  \BibitemOpen
  \bibfield  {author} {\bibinfo {author} {\bibfnamefont {D.~A.}\ \bibnamefont
  {Dikin}}, \bibinfo {author} {\bibfnamefont {M.}~\bibnamefont {Mehta}},
  \bibinfo {author} {\bibfnamefont {C.~W.}\ \bibnamefont {Bark}}, \bibinfo
  {author} {\bibfnamefont {C.~M.}\ \bibnamefont {Folkman}}, \bibinfo {author}
  {\bibfnamefont {C.~B.}\ \bibnamefont {Eom}}, \ and\ \bibinfo {author}
  {\bibfnamefont {V.}~\bibnamefont {Chandrasekhar}},\ }\href {\doibase
  10.1103/PhysRevLett.107.056802} {\bibfield  {journal} {\bibinfo  {journal}
  {Phys. Rev. Lett.}\ }\textbf {\bibinfo {volume} {107}},\ \bibinfo {pages}
  {056802} (\bibinfo {year} {2011})}\BibitemShut {NoStop}%
\bibitem [{\citenamefont {Bert}\ \emph {et~al.}(2011)\citenamefont {Bert},
  \citenamefont {Kalisky}, \citenamefont {Bell}, \citenamefont {Kim},
  \citenamefont {Hikita}, \citenamefont {Hwang},\ and\ \citenamefont
  {Moler}}]{Bert:natp11}%
  \BibitemOpen
  \bibfield  {author} {\bibinfo {author} {\bibfnamefont {J.~A.}\ \bibnamefont
  {Bert}}, \bibinfo {author} {\bibfnamefont {B.}~\bibnamefont {Kalisky}},
  \bibinfo {author} {\bibfnamefont {C.}~\bibnamefont {Bell}}, \bibinfo {author}
  {\bibfnamefont {M.}~\bibnamefont {Kim}}, \bibinfo {author} {\bibfnamefont
  {Y.}~\bibnamefont {Hikita}}, \bibinfo {author} {\bibfnamefont {H.~Y.}\
  \bibnamefont {Hwang}}, \ and\ \bibinfo {author} {\bibfnamefont {K.~A.}\
  \bibnamefont {Moler}},\ }\href {\doibase 10.1038/nphys2079} {\bibfield
  {journal} {\bibinfo  {journal} {Nature Physics}\ }\textbf {\bibinfo {volume}
  {7}},\ \bibinfo {pages} {767} (\bibinfo {year} {2011})}\BibitemShut {NoStop}%
\bibitem [{\citenamefont {Li}\ \emph {et~al.}(2011)\citenamefont {Li},
  \citenamefont {Richter}, \citenamefont {Mannhart},\ and\ \citenamefont
  {Ashoori}}]{Li:natp11}%
  \BibitemOpen
  \bibfield  {author} {\bibinfo {author} {\bibfnamefont {L.}~\bibnamefont
  {Li}}, \bibinfo {author} {\bibfnamefont {C.}~\bibnamefont {Richter}},
  \bibinfo {author} {\bibfnamefont {J.}~\bibnamefont {Mannhart}}, \ and\
  \bibinfo {author} {\bibfnamefont {R.~C.}\ \bibnamefont {Ashoori}},\ }\href
  {\doibase 10.1038/nphys2080} {\bibfield  {journal} {\bibinfo  {journal}
  {Nature Physics}\ }\textbf {\bibinfo {volume} {7}},\ \bibinfo {pages} {762}
  (\bibinfo {year} {2011})}\BibitemShut {NoStop}%
\bibitem [{\citenamefont {Schooley}\ \emph {et~al.}(1964)\citenamefont
  {Schooley}, \citenamefont {Hosler},\ and\ \citenamefont
  {Cohen}}]{Schooley:prl64}%
  \BibitemOpen
  \bibfield  {author} {\bibinfo {author} {\bibfnamefont {J.~F.}\ \bibnamefont
  {Schooley}}, \bibinfo {author} {\bibfnamefont {W.~R.}\ \bibnamefont
  {Hosler}}, \ and\ \bibinfo {author} {\bibfnamefont {M.~L.}\ \bibnamefont
  {Cohen}},\ }\href {\doibase 10.1103/PhysRevLett.12.474} {\bibfield  {journal}
  {\bibinfo  {journal} {Phys. Rev. Lett.}\ }\textbf {\bibinfo {volume} {12}},\
  \bibinfo {pages} {474} (\bibinfo {year} {1964})}\BibitemShut {NoStop}%
\bibitem [{\citenamefont {Brinkman}\ \emph {et~al.}(2007)\citenamefont
  {Brinkman}, \citenamefont {Huijben}, \citenamefont {van Zalk}, \citenamefont
  {Huijben}, \citenamefont {Zeitler}, \citenamefont {Maan}, \citenamefont {{van
  der Wiel}}, \citenamefont {Rijnders}, \citenamefont {Blank},\ and\
  \citenamefont {Hilgenkamp}}]{Brinkman:natm07}%
  \BibitemOpen
  \bibfield  {author} {\bibinfo {author} {\bibfnamefont {A.}~\bibnamefont
  {Brinkman}}, \bibinfo {author} {\bibfnamefont {M.}~\bibnamefont {Huijben}},
  \bibinfo {author} {\bibfnamefont {M.}~\bibnamefont {van Zalk}}, \bibinfo
  {author} {\bibfnamefont {J.}~\bibnamefont {Huijben}}, \bibinfo {author}
  {\bibfnamefont {U.}~\bibnamefont {Zeitler}}, \bibinfo {author} {\bibfnamefont
  {J.~C.}\ \bibnamefont {Maan}}, \bibinfo {author} {\bibfnamefont {W.~G.}\
  \bibnamefont {{van der Wiel}}}, \bibinfo {author} {\bibfnamefont
  {G.}~\bibnamefont {Rijnders}}, \bibinfo {author} {\bibfnamefont {D.~H.~A.}\
  \bibnamefont {Blank}}, \ and\ \bibinfo {author} {\bibfnamefont
  {H.}~\bibnamefont {Hilgenkamp}},\ }\href {\doibase 10.1038/nmat1931}
  {\bibfield  {journal} {\bibinfo  {journal} {Nature Materials}\ }\textbf
  {\bibinfo {volume} {6}},\ \bibinfo {pages} {493} (\bibinfo {year}
  {2007})}\BibitemShut {NoStop}%
\bibitem [{\citenamefont {Ariando}\ \emph {et~al.}(2011)\citenamefont
  {Ariando}, \citenamefont {Wang}, \citenamefont {Baskaran}, \citenamefont
  {Liu}, \citenamefont {Huijben}, \citenamefont {Yi}, \citenamefont {Annadi},
  \citenamefont {Barman}, \citenamefont {Rusydi}, \citenamefont {Dhar},
  \citenamefont {Feng}, \citenamefont {Ding}, \citenamefont {Hilgenkamp},\ and\
  \citenamefont {Venkatesan}}]{Ariando:natc11}%
  \BibitemOpen
  \bibfield  {author} {\bibinfo {author} {\bibnamefont {Ariando}}, \bibinfo
  {author} {\bibfnamefont {X.}~\bibnamefont {Wang}}, \bibinfo {author}
  {\bibfnamefont {G.}~\bibnamefont {Baskaran}}, \bibinfo {author}
  {\bibfnamefont {Z.~Q.}\ \bibnamefont {Liu}}, \bibinfo {author} {\bibfnamefont
  {J.}~\bibnamefont {Huijben}}, \bibinfo {author} {\bibfnamefont {J.~B.}\
  \bibnamefont {Yi}}, \bibinfo {author} {\bibfnamefont {A.}~\bibnamefont
  {Annadi}}, \bibinfo {author} {\bibfnamefont {A.~R.}\ \bibnamefont {Barman}},
  \bibinfo {author} {\bibfnamefont {A.}~\bibnamefont {Rusydi}}, \bibinfo
  {author} {\bibfnamefont {S.}~\bibnamefont {Dhar}}, \bibinfo {author}
  {\bibfnamefont {Y.~P.}\ \bibnamefont {Feng}}, \bibinfo {author}
  {\bibfnamefont {J.}~\bibnamefont {Ding}}, \bibinfo {author} {\bibfnamefont
  {H.}~\bibnamefont {Hilgenkamp}}, \ and\ \bibinfo {author} {\bibfnamefont
  {T.}~\bibnamefont {Venkatesan}},\ }\href {\doibase 10.1038/ncomms1192}
  {\bibfield  {journal} {\bibinfo  {journal} {Nature Communications}\ }\textbf
  {\bibinfo {volume} {2}},\ \bibinfo {pages} {188} (\bibinfo {year}
  {2011})}\BibitemShut {NoStop}%
\bibitem [{\citenamefont {Fitzsimmons}\ \emph {et~al.}(2011)\citenamefont
  {Fitzsimmons}, \citenamefont {Hengartner}, \citenamefont {Singh},
  \citenamefont {Zhernenkov}, \citenamefont {Bruno}, \citenamefont
  {Santamaria}, \citenamefont {Brinkman}, \citenamefont {Huijben},
  \citenamefont {Molegraaf}, \citenamefont {{de la Venta}},\ and\ \citenamefont
  {Schuller}}]{Fitzsimmons:prl11}%
  \BibitemOpen
  \bibfield  {author} {\bibinfo {author} {\bibfnamefont {M.~R.}\ \bibnamefont
  {Fitzsimmons}}, \bibinfo {author} {\bibfnamefont {N.~W.}\ \bibnamefont
  {Hengartner}}, \bibinfo {author} {\bibfnamefont {S.}~\bibnamefont {Singh}},
  \bibinfo {author} {\bibfnamefont {M.}~\bibnamefont {Zhernenkov}}, \bibinfo
  {author} {\bibfnamefont {F.~Y.}\ \bibnamefont {Bruno}}, \bibinfo {author}
  {\bibfnamefont {J.}~\bibnamefont {Santamaria}}, \bibinfo {author}
  {\bibfnamefont {A.}~\bibnamefont {Brinkman}}, \bibinfo {author}
  {\bibfnamefont {M.}~\bibnamefont {Huijben}}, \bibinfo {author} {\bibfnamefont
  {H.~J.~A.}\ \bibnamefont {Molegraaf}}, \bibinfo {author} {\bibfnamefont
  {J.}~\bibnamefont {{de la Venta}}}, \ and\ \bibinfo {author} {\bibfnamefont
  {I.~K.}\ \bibnamefont {Schuller}},\ }\href {\doibase
  10.1103/PhysRevLett.107.217201} {\bibfield  {journal} {\bibinfo  {journal}
  {Phys. Rev. Lett.}\ }\textbf {\bibinfo {volume} {107}},\ \bibinfo {pages}
  {217201} (\bibinfo {year} {2011})}\BibitemShut {NoStop}%
\bibitem [{\citenamefont {Kalisky}\ \emph {et~al.}(2012)\citenamefont
  {Kalisky}, \citenamefont {Bert}, \citenamefont {Klopfer}, \citenamefont
  {Bell}, \citenamefont {Sato}, \citenamefont {Hosoda}, \citenamefont {Hikita},
  \citenamefont {Hwang},\ and\ \citenamefont {Moler}}]{Kalisky:natc12}%
  \BibitemOpen
  \bibfield  {author} {\bibinfo {author} {\bibfnamefont {B.}~\bibnamefont
  {Kalisky}}, \bibinfo {author} {\bibfnamefont {J.~A.}\ \bibnamefont {Bert}},
  \bibinfo {author} {\bibfnamefont {B.~B.}\ \bibnamefont {Klopfer}}, \bibinfo
  {author} {\bibfnamefont {C.}~\bibnamefont {Bell}}, \bibinfo {author}
  {\bibfnamefont {H.~K.}\ \bibnamefont {Sato}}, \bibinfo {author}
  {\bibfnamefont {M.}~\bibnamefont {Hosoda}}, \bibinfo {author} {\bibfnamefont
  {Y.}~\bibnamefont {Hikita}}, \bibinfo {author} {\bibfnamefont {H.~Y.}\
  \bibnamefont {Hwang}}, \ and\ \bibinfo {author} {\bibfnamefont {K.~A.}\
  \bibnamefont {Moler}},\ }\href {\doibase 10.1038/ncomms1931} {\bibfield
  {journal} {\bibinfo  {journal} {Nature Communications}\ }\textbf {\bibinfo
  {volume} {3}},\ \bibinfo {pages} {922} (\bibinfo {year} {2012})}\BibitemShut
  {NoStop}%
\bibitem [{\citenamefont {Salman}\ \emph {et~al.}(2012)\citenamefont {Salman},
  \citenamefont {Ofer}, \citenamefont {Radovic}, \citenamefont {Hao},
  \citenamefont {{Ben Shalom}}, \citenamefont {Chow}, \citenamefont {Dagan},
  \citenamefont {Hossain}, \citenamefont {Levy}, \citenamefont {MacFarlane},
  \citenamefont {Morris}, \citenamefont {Patthey}, \citenamefont {Pearson},
  \citenamefont {Saadaoui}, \citenamefont {Schmitt}, \citenamefont {Wang},\
  and\ \citenamefont {Kiefl}}]{Salman:prl12}%
  \BibitemOpen
  \bibfield  {author} {\bibinfo {author} {\bibfnamefont {Z.}~\bibnamefont
  {Salman}}, \bibinfo {author} {\bibfnamefont {O.}~\bibnamefont {Ofer}},
  \bibinfo {author} {\bibfnamefont {M.}~\bibnamefont {Radovic}}, \bibinfo
  {author} {\bibfnamefont {H.}~\bibnamefont {Hao}}, \bibinfo {author}
  {\bibfnamefont {M.}~\bibnamefont {{Ben Shalom}}}, \bibinfo {author}
  {\bibfnamefont {K.~H.}\ \bibnamefont {Chow}}, \bibinfo {author}
  {\bibfnamefont {Y.}~\bibnamefont {Dagan}}, \bibinfo {author} {\bibfnamefont
  {M.~D.}\ \bibnamefont {Hossain}}, \bibinfo {author} {\bibfnamefont
  {C.~D.~P.}\ \bibnamefont {Levy}}, \bibinfo {author} {\bibfnamefont {W.~A.}\
  \bibnamefont {MacFarlane}}, \bibinfo {author} {\bibfnamefont {G.~M.}\
  \bibnamefont {Morris}}, \bibinfo {author} {\bibfnamefont {L.}~\bibnamefont
  {Patthey}}, \bibinfo {author} {\bibfnamefont {M.~R.}\ \bibnamefont
  {Pearson}}, \bibinfo {author} {\bibfnamefont {H.}~\bibnamefont {Saadaoui}},
  \bibinfo {author} {\bibfnamefont {T.}~\bibnamefont {Schmitt}}, \bibinfo
  {author} {\bibfnamefont {D.}~\bibnamefont {Wang}}, \ and\ \bibinfo {author}
  {\bibfnamefont {R.~F.}\ \bibnamefont {Kiefl}},\ }\href {\doibase
  10.1103/PhysRevLett.109.257207} {\bibfield  {journal} {\bibinfo  {journal}
  {Phys. Rev. Lett.}\ }\textbf {\bibinfo {volume} {109}},\ \bibinfo {pages}
  {257207} (\bibinfo {year} {2012})}\BibitemShut {NoStop}%
\bibitem [{\citenamefont {Lee}\ \emph {et~al.}(2013)\citenamefont {Lee},
  \citenamefont {Xie}, \citenamefont {Sato}, \citenamefont {Bell},
  \citenamefont {Hikita}, \citenamefont {Hwang},\ and\ \citenamefont
  {Kao}}]{Lee:natm13}%
  \BibitemOpen
  \bibfield  {author} {\bibinfo {author} {\bibfnamefont {J.-S.}\ \bibnamefont
  {Lee}}, \bibinfo {author} {\bibfnamefont {Y.~W.}\ \bibnamefont {Xie}},
  \bibinfo {author} {\bibfnamefont {H.~K.}\ \bibnamefont {Sato}}, \bibinfo
  {author} {\bibfnamefont {C.}~\bibnamefont {Bell}}, \bibinfo {author}
  {\bibfnamefont {Y.}~\bibnamefont {Hikita}}, \bibinfo {author} {\bibfnamefont
  {H.~Y.}\ \bibnamefont {Hwang}}, \ and\ \bibinfo {author} {\bibfnamefont
  {C.-C.}\ \bibnamefont {Kao}},\ }\href {\doibase 10.1038/nmat3674} {\bibfield
  {journal} {\bibinfo  {journal} {Nature Materials}\ }\textbf {\bibinfo
  {volume} {12}},\ \bibinfo {pages} {703} (\bibinfo {year} {2013})}\BibitemShut
  {NoStop}%
\bibitem [{\citenamefont {Pentcheva}\ and\ \citenamefont
  {Pickett}(2006)}]{Pentcheva:prb06}%
  \BibitemOpen
  \bibfield  {author} {\bibinfo {author} {\bibfnamefont {R.}~\bibnamefont
  {Pentcheva}}\ and\ \bibinfo {author} {\bibfnamefont {W.~E.}\ \bibnamefont
  {Pickett}},\ }\href {\doibase 10.1103/PhysRevB.74.035112} {\bibfield
  {journal} {\bibinfo  {journal} {Phys. Rev. B}\ }\textbf {\bibinfo {volume}
  {74}},\ \bibinfo {pages} {035112} (\bibinfo {year} {2006})}\BibitemShut
  {NoStop}%
\bibitem [{\citenamefont {Janicka}\ \emph {et~al.}(2008)\citenamefont
  {Janicka}, \citenamefont {Velev},\ and\ \citenamefont
  {Tsymbal}}]{Janicka:jap08}%
  \BibitemOpen
  \bibfield  {author} {\bibinfo {author} {\bibfnamefont {K.}~\bibnamefont
  {Janicka}}, \bibinfo {author} {\bibfnamefont {J.~P.}\ \bibnamefont {Velev}},
  \ and\ \bibinfo {author} {\bibfnamefont {E.~Y.}\ \bibnamefont {Tsymbal}},\
  }\href {\doibase 10.1063/1.2829244} {\bibfield  {journal} {\bibinfo
  {journal} {J. Appl. Phys.}\ }\textbf {\bibinfo {volume} {103}},\ \bibinfo
  {pages} {078508} (\bibinfo {year} {2008})}\BibitemShut {NoStop}%
\bibitem [{\citenamefont {Zhong}\ and\ \citenamefont
  {Kelly}(2008)}]{Zhong:epl08}%
  \BibitemOpen
  \bibfield  {author} {\bibinfo {author} {\bibfnamefont {Z.}~\bibnamefont
  {Zhong}}\ and\ \bibinfo {author} {\bibfnamefont {P.~J.}\ \bibnamefont
  {Kelly}},\ }\href {\doibase 10.1209/0295-5075/84/27001} {\bibfield  {journal}
  {\bibinfo  {journal} {Europhys. Lett.}\ }\textbf {\bibinfo {volume} {84}},\
  \bibinfo {pages} {27001} (\bibinfo {year} {2008})}\BibitemShut {NoStop}%
\bibitem [{\citenamefont {Li}\ \emph {et~al.}(2013)\citenamefont {Li},
  \citenamefont {Beltr{\'{a}}n},\ and\ \citenamefont {{Carmen
  Mu{\~{n}}oz}}}]{Li:prb13}%
  \BibitemOpen
  \bibfield  {author} {\bibinfo {author} {\bibfnamefont {J.}~\bibnamefont
  {Li}}, \bibinfo {author} {\bibfnamefont {J.}~\bibnamefont {Beltr{\'{a}}n}}, \
  and\ \bibinfo {author} {\bibfnamefont {M.}~\bibnamefont {{Carmen
  Mu{\~{n}}oz}}},\ }\href {\doibase 10.1103/PhysRevB.87.075411} {\bibfield
  {journal} {\bibinfo  {journal} {Phys. Rev. B}\ }\textbf {\bibinfo {volume}
  {87}},\ \bibinfo {pages} {075411} (\bibinfo {year} {2013})}\BibitemShut
  {NoStop}%
\bibitem [{\citenamefont {Gunnarsson}(1976)}]{Gunnarsson:jpf76}%
  \BibitemOpen
  \bibfield  {author} {\bibinfo {author} {\bibfnamefont {O.}~\bibnamefont
  {Gunnarsson}},\ }\href {\doibase 10.1088/0305-4608/6/4/018} {\bibfield
  {journal} {\bibinfo  {journal} {J. Phys. F: Met. Phys.}\ }\textbf {\bibinfo
  {volume} {6}},\ \bibinfo {pages} {587} (\bibinfo {year} {1976})}\BibitemShut
  {NoStop}%
\bibitem [{\citenamefont {Poulsen}\ \emph {et~al.}(1976)\citenamefont
  {Poulsen}, \citenamefont {Kollar},\ and\ \citenamefont
  {Andersen}}]{Poulsen:jpf76}%
  \BibitemOpen
  \bibfield  {author} {\bibinfo {author} {\bibfnamefont {U.~K.}\ \bibnamefont
  {Poulsen}}, \bibinfo {author} {\bibfnamefont {J.}~\bibnamefont {Kollar}}, \
  and\ \bibinfo {author} {\bibfnamefont {O.~K.}\ \bibnamefont {Andersen}},\
  }\href {\doibase 10.1088/0305-4608/6/9/002} {\bibfield  {journal} {\bibinfo
  {journal} {J. Phys. F: Met. Phys.}\ }\textbf {\bibinfo {volume} {6}},\
  \bibinfo {pages} {L241} (\bibinfo {year} {1976})}\BibitemShut {NoStop}%
\bibitem [{\citenamefont {Janak}(1977)}]{Janak:prb77}%
  \BibitemOpen
  \bibfield  {author} {\bibinfo {author} {\bibfnamefont {J.~F.}\ \bibnamefont
  {Janak}},\ }\href {\doibase 10.1103/PhysRevB.16.255} {\bibfield  {journal}
  {\bibinfo  {journal} {Phys. Rev. B}\ }\textbf {\bibinfo {volume} {16}},\
  \bibinfo {pages} {255} (\bibinfo {year} {1977})}\BibitemShut {NoStop}%
\bibitem [{\citenamefont {Balamurugan}\ \emph {et~al.}(2010)\citenamefont
  {Balamurugan}, \citenamefont {Rodewald}, \citenamefont {Harmening},
  \citenamefont {{van W{\"{u}}llen}}, \citenamefont {Mohr}, \citenamefont
  {Deters}, \citenamefont {Eckert},\ and\ \citenamefont
  {P{\"{o}}ttgen}}]{Balamurugan:znb10}%
  \BibitemOpen
  \bibfield  {author} {\bibinfo {author} {\bibfnamefont {S.}~\bibnamefont
  {Balamurugan}}, \bibinfo {author} {\bibfnamefont {U.~C.}\ \bibnamefont
  {Rodewald}}, \bibinfo {author} {\bibfnamefont {T.}~\bibnamefont {Harmening}},
  \bibinfo {author} {\bibfnamefont {L.}~\bibnamefont {{van W{\"{u}}llen}}},
  \bibinfo {author} {\bibfnamefont {D.}~\bibnamefont {Mohr}}, \bibinfo {author}
  {\bibfnamefont {H.}~\bibnamefont {Deters}}, \bibinfo {author} {\bibfnamefont
  {H.}~\bibnamefont {Eckert}}, \ and\ \bibinfo {author} {\bibfnamefont
  {R.}~\bibnamefont {P{\"{o}}ttgen}},\ }\href@noop {} {\bibfield  {journal}
  {\bibinfo  {journal} {Z. Naturforsch. B: Chem. Sci.}\ }\textbf {\bibinfo
  {volume} {65b}},\ \bibinfo {pages} {1199} (\bibinfo {year}
  {2010})}\BibitemShut {NoStop}%
\bibitem [{\citenamefont {Howard}\ and\ \citenamefont
  {Stokes}(1998)}]{Howard:acsb98}%
  \BibitemOpen
  \bibfield  {author} {\bibinfo {author} {\bibfnamefont {C.~J.}\ \bibnamefont
  {Howard}}\ and\ \bibinfo {author} {\bibfnamefont {H.~T.}\ \bibnamefont
  {Stokes}},\ }\href {\doibase 10.1107/S0108768198004200} {\bibfield  {journal}
  {\bibinfo  {journal} {Acta Crystallogr. Sect. B: Struct. Sci.}\ }\textbf
  {\bibinfo {volume} {B54}},\ \bibinfo {pages} {782} (\bibinfo {year}
  {1998})}\BibitemShut {NoStop}%
\bibitem [{\citenamefont {Sasaki}\ \emph {et~al.}(1987)\citenamefont {Sasaki},
  \citenamefont {Prewitt}, \citenamefont {Bass},\ and\ \citenamefont
  {Schulze}}]{Sasaki:acsc87}%
  \BibitemOpen
  \bibfield  {author} {\bibinfo {author} {\bibfnamefont {S.}~\bibnamefont
  {Sasaki}}, \bibinfo {author} {\bibfnamefont {C.~T.}\ \bibnamefont {Prewitt}},
  \bibinfo {author} {\bibfnamefont {J.~D.}\ \bibnamefont {Bass}}, \ and\
  \bibinfo {author} {\bibfnamefont {W.~A.}\ \bibnamefont {Schulze}},\ }\href
  {\doibase 10.1107/S0108270187090620} {\bibfield  {journal} {\bibinfo
  {journal} {Acta Cryst.}\ }\textbf {\bibinfo {volume} {C43}},\ \bibinfo
  {pages} {1668} (\bibinfo {year} {1987})}\BibitemShut {NoStop}%
\bibitem [{\citenamefont {Bl\"{o}chl}(1994)}]{paw}%
  \BibitemOpen
  \bibfield  {author} {\bibinfo {author} {\bibfnamefont {P.~E.}\ \bibnamefont
  {Bl\"{o}chl}},\ }\href {\doibase 10.1103/PhysRevB.50.17953} {\bibfield
  {journal} {\bibinfo  {journal} {Phys. Rev. B}\ }\textbf {\bibinfo {volume}
  {50}},\ \bibinfo {pages} {17953} (\bibinfo {year} {1994})}\BibitemShut
  {NoStop}%
\bibitem [{\citenamefont {Kresse}\ and\ \citenamefont
  {Hafner}(1993)}]{Kresse:prb93}%
  \BibitemOpen
  \bibfield  {author} {\bibinfo {author} {\bibfnamefont {G.}~\bibnamefont
  {Kresse}}\ and\ \bibinfo {author} {\bibfnamefont {J.}~\bibnamefont
  {Hafner}},\ }\href {\doibase 10.1103/PhysRevB.47.558} {\bibfield  {journal}
  {\bibinfo  {journal} {Phys. Rev. B}\ }\textbf {\bibinfo {volume} {47}},\
  \bibinfo {pages} {558} (\bibinfo {year} {1993})}\BibitemShut {NoStop}%
\bibitem [{\citenamefont {Kresse}\ and\ \citenamefont
  {Furthm{\"{u}}ller}(1996)}]{Kresse:prb96}%
  \BibitemOpen
  \bibfield  {author} {\bibinfo {author} {\bibfnamefont {G.}~\bibnamefont
  {Kresse}}\ and\ \bibinfo {author} {\bibfnamefont {J.}~\bibnamefont
  {Furthm{\"{u}}ller}},\ }\href {\doibase 10.1103/PhysRevB.54.11169} {\bibfield
   {journal} {\bibinfo  {journal} {Phys. Rev. B}\ }\textbf {\bibinfo {volume}
  {54}},\ \bibinfo {pages} {11169} (\bibinfo {year} {1996})}\BibitemShut
  {NoStop}%
\bibitem [{\citenamefont {Perdew}\ and\ \citenamefont
  {Zunger}(1981)}]{Perdew:prb81}%
  \BibitemOpen
  \bibfield  {author} {\bibinfo {author} {\bibfnamefont {J.~P.}\ \bibnamefont
  {Perdew}}\ and\ \bibinfo {author} {\bibfnamefont {A.}~\bibnamefont
  {Zunger}},\ }\href {\doibase 10.1103/PhysRevB.23.5048} {\bibfield  {journal}
  {\bibinfo  {journal} {Phys. Rev. B}\ }\textbf {\bibinfo {volume} {23}},\
  \bibinfo {pages} {5048} (\bibinfo {year} {1981})}\BibitemShut {NoStop}%
\bibitem [{\citenamefont {Dudarev}\ \emph {et~al.}(1998)\citenamefont
  {Dudarev}, \citenamefont {Botton}, \citenamefont {Savrasov}, \citenamefont
  {Humphreys},\ and\ \citenamefont {Sutton}}]{Dudarev:prb98}%
  \BibitemOpen
  \bibfield  {author} {\bibinfo {author} {\bibfnamefont {S.~L.}\ \bibnamefont
  {Dudarev}}, \bibinfo {author} {\bibfnamefont {G.~A.}\ \bibnamefont {Botton}},
  \bibinfo {author} {\bibfnamefont {S.~Y.}\ \bibnamefont {Savrasov}}, \bibinfo
  {author} {\bibfnamefont {C.~J.}\ \bibnamefont {Humphreys}}, \ and\ \bibinfo
  {author} {\bibfnamefont {A.~P.}\ \bibnamefont {Sutton}},\ }\href {\doibase
  10.1103/PhysRevB.57.150} {\bibfield  {journal} {\bibinfo  {journal} {Phys.
  Rev. B}\ }\textbf {\bibinfo {volume} {57}},\ \bibinfo {pages} {1505}
  (\bibinfo {year} {1998})}\BibitemShut {NoStop}%
\bibitem [{\citenamefont {Ganguli}\ \emph {et~al.}()\citenamefont {Ganguli},
  \citenamefont {Zhong},\ and\ \citenamefont {Kelly}}]{Ganguli:d1Oxide}%
  \BibitemOpen
  \bibfield  {author} {\bibinfo {author} {\bibfnamefont {N.}~\bibnamefont
  {Ganguli}}, \bibinfo {author} {\bibfnamefont {Z.}~\bibnamefont {Zhong}}, \
  and\ \bibinfo {author} {\bibfnamefont {P.~J.}\ \bibnamefont {Kelly}},\
  }\href@noop {} {\bibinfo  {journal} {to be published}\ }\BibitemShut
  {NoStop}%
\bibitem [{\citenamefont {Okatov}\ \emph {et~al.}(2005)\citenamefont {Okatov},
  \citenamefont {Poteryaev},\ and\ \citenamefont
  {Lichtenstein}}]{Okatov:epl05}%
  \BibitemOpen
\bibfield  {journal} {  }\bibfield  {author} {\bibinfo {author} {\bibfnamefont
  {S.}~\bibnamefont {Okatov}}, \bibinfo {author} {\bibfnamefont
  {A.}~\bibnamefont {Poteryaev}}, \ and\ \bibinfo {author} {\bibfnamefont
  {A.}~\bibnamefont {Lichtenstein}},\ }\href {\doibase
  10.1209/epl/i2004-10513-x} {\bibfield  {journal} {\bibinfo  {journal}
  {Europhys. Lett.}\ }\textbf {\bibinfo {volume} {70}},\ \bibinfo {pages} {499}
  (\bibinfo {year} {2005})}\BibitemShut {NoStop}%
\bibitem [{Note1()}]{Note1}%
  \BibitemOpen
  \bibinfo {note} {We find an $a^+b^-c^-$ tilt system for the TiO$_6$ octahedra
  near the LAO$|$CTO interface as opposed to an $a^+b^-b^-$ tilt system for
  bulk CaTiO$_3$ \cite {Woodward:acsb97b}.}\BibitemShut {Stop}%
\bibitem [{\citenamefont {Andersen}\ \emph {et~al.}(1985)\citenamefont
  {Andersen}, \citenamefont {Jepsen},\ and\ \citenamefont
  {Gl{\"{o}}tzel}}]{Andersen:85}%
  \BibitemOpen
  \bibfield  {author} {\bibinfo {author} {\bibfnamefont {O.~K.}\ \bibnamefont
  {Andersen}}, \bibinfo {author} {\bibfnamefont {O.}~\bibnamefont {Jepsen}}, \
  and\ \bibinfo {author} {\bibfnamefont {D.}~\bibnamefont {Gl{\"{o}}tzel}},\
  }in\ \href@noop {} {\emph {\bibinfo {booktitle} {Highlights of Condensed
  Matter Theory}}},\ \bibinfo {series and number} {International School of
  Physics `Enrico Fermi', Varenna, Italy},\ \bibinfo {editor} {edited by\
  \bibinfo {editor} {\bibfnamefont {F.}~\bibnamefont {Bassani}}, \bibinfo
  {editor} {\bibfnamefont {F.}~\bibnamefont {Fumi}}, \ and\ \bibinfo {editor}
  {\bibfnamefont {M.~P.}\ \bibnamefont {Tosi}}}\ (\bibinfo  {publisher}
  {North-Holland},\ \bibinfo {address} {Amsterdam},\ \bibinfo {year} {1985})\
  pp.\ \bibinfo {pages} {59--176}\BibitemShut {NoStop}%
\bibitem [{\citenamefont {Popovi{\'{c}}}\ \emph {et~al.}(2008)\citenamefont
  {Popovi{\'{c}}}, \citenamefont {Satpathy},\ and\ \citenamefont
  {Martin}}]{Popovic:prl08}%
  \BibitemOpen
  \bibfield  {author} {\bibinfo {author} {\bibfnamefont {Z.~S.}\ \bibnamefont
  {Popovi{\'{c}}}}, \bibinfo {author} {\bibfnamefont {S.}~\bibnamefont
  {Satpathy}}, \ and\ \bibinfo {author} {\bibfnamefont {R.~M.}\ \bibnamefont
  {Martin}},\ }\href {\doibase 10.1103/PhysRevLett.101.256801} {\bibfield
  {journal} {\bibinfo  {journal} {Phys. Rev. Lett.}\ }\textbf {\bibinfo
  {volume} {101}},\ \bibinfo {pages} {256801} (\bibinfo {year}
  {2008})}\BibitemShut {NoStop}%
\bibitem [{\citenamefont {Delugas}\ \emph {et~al.}(2011)\citenamefont
  {Delugas}, \citenamefont {Filippetti}, \citenamefont {Fiorentini},
  \citenamefont {Bilc}, \citenamefont {Fontaine},\ and\ \citenamefont
  {Ghosez}}]{Delugas:prl11}%
  \BibitemOpen
  \bibfield  {author} {\bibinfo {author} {\bibfnamefont {P.}~\bibnamefont
  {Delugas}}, \bibinfo {author} {\bibfnamefont {A.}~\bibnamefont {Filippetti}},
  \bibinfo {author} {\bibfnamefont {V.}~\bibnamefont {Fiorentini}}, \bibinfo
  {author} {\bibfnamefont {D.~I.}\ \bibnamefont {Bilc}}, \bibinfo {author}
  {\bibfnamefont {D.}~\bibnamefont {Fontaine}}, \ and\ \bibinfo {author}
  {\bibfnamefont {P.}~\bibnamefont {Ghosez}},\ }\href {\doibase
  10.1103/PhysRevLett.106.166807} {\bibfield  {journal} {\bibinfo  {journal}
  {Phys. Rev. Lett.}\ }\textbf {\bibinfo {volume} {106}},\ \bibinfo {pages}
  {166807} (\bibinfo {year} {2011})}\BibitemShut {NoStop}%
\bibitem [{Note2()}]{Note2}%
  \BibitemOpen
  \bibinfo {note} {For inhomogeneous systems, the Stoner criterion is at best
  indicative because of the arbitrariness in defining density of state
  projections and because the exchange polarization in complex materials more
  often than not does not have rigid-band character.}\BibitemShut {Stop}%
\bibitem [{\citenamefont {Mizokawa}\ and\ \citenamefont
  {Fujimori}(1995)}]{MizokawaPRB95}%
  \BibitemOpen
  \bibfield  {author} {\bibinfo {author} {\bibfnamefont {T.}~\bibnamefont
  {Mizokawa}}\ and\ \bibinfo {author} {\bibfnamefont {A.}~\bibnamefont
  {Fujimori}},\ }\href {\doibase 10.1103/PhysRevB.51.12880} {\bibfield
  {journal} {\bibinfo  {journal} {Phys. Rev. B}\ }\textbf {\bibinfo {volume}
  {51}},\ \bibinfo {pages} {12880} (\bibinfo {year} {1995})}\BibitemShut
  {NoStop}%
\bibitem [{\citenamefont {Nakagawa}\ \emph {et~al.}(2006)\citenamefont
  {Nakagawa}, \citenamefont {Hwang},\ and\ \citenamefont
  {Muller}}]{Nakagawa:natm06}%
  \BibitemOpen
  \bibfield  {author} {\bibinfo {author} {\bibfnamefont {N.}~\bibnamefont
  {Nakagawa}}, \bibinfo {author} {\bibfnamefont {H.~Y.}\ \bibnamefont {Hwang}},
  \ and\ \bibinfo {author} {\bibfnamefont {D.~A.}\ \bibnamefont {Muller}},\
  }\href {\doibase 10.1038/nmat1569} {\bibfield  {journal} {\bibinfo  {journal}
  {Nature Materials}\ }\textbf {\bibinfo {volume} {5}},\ \bibinfo {pages} {204}
  (\bibinfo {year} {2006})}\BibitemShut {NoStop}%
\bibitem [{\citenamefont {McCollam}\ \emph {et~al.}(2014)\citenamefont
  {McCollam}, \citenamefont {Wenderich}, \citenamefont {Kruize}, \citenamefont
  {Guduru}, \citenamefont {Molegraaf}, \citenamefont {Huijben}, \citenamefont
  {Koster}, \citenamefont {Blank}, \citenamefont {Rijnders}, \citenamefont
  {Brinkman}, \citenamefont {Hilgenkamp}, \citenamefont {Zeitler},\ and\
  \citenamefont {Maan}}]{McCollam:aplm14}%
  \BibitemOpen
  \bibfield  {author} {\bibinfo {author} {\bibfnamefont {A.}~\bibnamefont
  {McCollam}}, \bibinfo {author} {\bibfnamefont {S.}~\bibnamefont {Wenderich}},
  \bibinfo {author} {\bibfnamefont {M.~K.}\ \bibnamefont {Kruize}}, \bibinfo
  {author} {\bibfnamefont {V.~K.}\ \bibnamefont {Guduru}}, \bibinfo {author}
  {\bibfnamefont {H.~J.~A.}\ \bibnamefont {Molegraaf}}, \bibinfo {author}
  {\bibfnamefont {M.}~\bibnamefont {Huijben}}, \bibinfo {author} {\bibfnamefont
  {G.}~\bibnamefont {Koster}}, \bibinfo {author} {\bibfnamefont {D.~H.~A.}\
  \bibnamefont {Blank}}, \bibinfo {author} {\bibfnamefont {G.}~\bibnamefont
  {Rijnders}}, \bibinfo {author} {\bibfnamefont {A.}~\bibnamefont {Brinkman}},
  \bibinfo {author} {\bibfnamefont {H.}~\bibnamefont {Hilgenkamp}}, \bibinfo
  {author} {\bibfnamefont {U.}~\bibnamefont {Zeitler}}, \ and\ \bibinfo
  {author} {\bibfnamefont {J.~C.}\ \bibnamefont {Maan}},\ }\href {\doibase
  10.1063/1.4863786} {\bibfield  {journal} {\bibinfo  {journal} {APL
  Materials}\ }\textbf {\bibinfo {volume} {2}},\ \bibinfo {pages} {022102}
  (\bibinfo {year} {2014})}\BibitemShut {NoStop}%
\bibitem [{\citenamefont {Konaka}\ \emph {et~al.}(1991)\citenamefont {Konaka},
  \citenamefont {Sato}, \citenamefont {Asano},\ and\ \citenamefont
  {Kubo}}]{Konaka:js91}%
  \BibitemOpen
  \bibfield  {author} {\bibinfo {author} {\bibfnamefont {T.}~\bibnamefont
  {Konaka}}, \bibinfo {author} {\bibfnamefont {M.}~\bibnamefont {Sato}},
  \bibinfo {author} {\bibfnamefont {H.}~\bibnamefont {Asano}}, \ and\ \bibinfo
  {author} {\bibfnamefont {S.}~\bibnamefont {Kubo}},\ }\href {\doibase
  10.1007/BF00618150} {\bibfield  {journal} {\bibinfo  {journal} {J. of
  Superconductivity}\ }\textbf {\bibinfo {volume} {4}},\ \bibinfo {pages} {283}
  (\bibinfo {year} {1991})}\BibitemShut {NoStop}%
\bibitem [{\citenamefont {Ueda}\ \emph {et~al.}(1998)\citenamefont {Ueda},
  \citenamefont {Yangi}, \citenamefont {Noshiro}, \citenamefont {Hosono},\ and\
  \citenamefont {Kawazoe}}]{Ueda:jpcm98}%
  \BibitemOpen
  \bibfield  {author} {\bibinfo {author} {\bibfnamefont {K.}~\bibnamefont
  {Ueda}}, \bibinfo {author} {\bibfnamefont {H.}~\bibnamefont {Yangi}},
  \bibinfo {author} {\bibfnamefont {R.}~\bibnamefont {Noshiro}}, \bibinfo
  {author} {\bibfnamefont {H.}~\bibnamefont {Hosono}}, \ and\ \bibinfo {author}
  {\bibfnamefont {H.}~\bibnamefont {Kawazoe}},\ }\href {\doibase
  10.1088/0953-8984/10/16/018} {\bibfield  {journal} {\bibinfo  {journal} {J.
  Phys.: Condens. Matter}\ }\textbf {\bibinfo {volume} {10}},\ \bibinfo {pages}
  {3669} (\bibinfo {year} {1998})}\BibitemShut {NoStop}%
\bibitem [{Note3()}]{Note3}%
  \BibitemOpen
  \bibinfo {note} {We found no significant change in $\sigma ^\protect \text
  {t} / \sigma $ as a function of LAO thickness for $\left (\protect
  \mathaccentV {hat}05E{a}^\protect \text {CTO}/{2 \epsilon ^\protect \text
  {LAO}} = 0.89 \right ) \geq K^\protect \text {IF} \geq \left (\protect
  \mathaccentV {hat}05E{a}^\protect \text {CTO}/{2 \epsilon ^\protect \text
  {CTO}} = 0.11 \right )$~F$^{-1}$. $\epsilon ^\protect \text {CTO}/\epsilon _0
  = 190$ at room temperature \cite {Linz:jcp58}.}\BibitemShut {Stop}%
\bibitem [{\citenamefont {Thiel}\ \emph {et~al.}(2006)\citenamefont {Thiel},
  \citenamefont {Hammerl}, \citenamefont {Schmehl}, \citenamefont {Schneider},\
  and\ \citenamefont {Mannhart}}]{Thiel:sc06}%
  \BibitemOpen
  \bibfield  {author} {\bibinfo {author} {\bibfnamefont {S.}~\bibnamefont
  {Thiel}}, \bibinfo {author} {\bibfnamefont {G.}~\bibnamefont {Hammerl}},
  \bibinfo {author} {\bibfnamefont {A.}~\bibnamefont {Schmehl}}, \bibinfo
  {author} {\bibfnamefont {C.~W.}\ \bibnamefont {Schneider}}, \ and\ \bibinfo
  {author} {\bibfnamefont {J.}~\bibnamefont {Mannhart}},\ }\href {\doibase
  10.1126/science.1131091} {\bibfield  {journal} {\bibinfo  {journal}
  {Science}\ }\textbf {\bibinfo {volume} {313}},\ \bibinfo {pages} {1942}
  (\bibinfo {year} {2006})}\BibitemShut {NoStop}%
\bibitem [{\citenamefont {Xie}\ \emph {et~al.}(2011)\citenamefont {Xie},
  \citenamefont {Hikita}, \citenamefont {Bell},\ and\ \citenamefont
  {Hwang}}]{Xie:natc11}%
  \BibitemOpen
  \bibfield  {author} {\bibinfo {author} {\bibfnamefont {Y.}~\bibnamefont
  {Xie}}, \bibinfo {author} {\bibfnamefont {Y.}~\bibnamefont {Hikita}},
  \bibinfo {author} {\bibfnamefont {C.}~\bibnamefont {Bell}}, \ and\ \bibinfo
  {author} {\bibfnamefont {H.~Y.}\ \bibnamefont {Hwang}},\ }\href {\doibase
  10.1038/ncomms1501} {\bibfield  {journal} {\bibinfo  {journal} {Nature
  Communications}\ }\textbf {\bibinfo {volume} {2}},\ \bibinfo {pages} {494}
  (\bibinfo {year} {2011})}\BibitemShut {NoStop}%
\bibitem [{\citenamefont {Bi}\ \emph {et~al.}()\citenamefont {Bi},
  \citenamefont {Huang}, \citenamefont {Bark}, \citenamefont {Ryu},
  \citenamefont {Eom},\ and\ \citenamefont {Levy}}]{Bi:arXiv13}%
  \BibitemOpen
  \bibfield  {author} {\bibinfo {author} {\bibfnamefont {F.}~\bibnamefont
  {Bi}}, \bibinfo {author} {\bibfnamefont {M.}~\bibnamefont {Huang}}, \bibinfo
  {author} {\bibfnamefont {C.-W.}\ \bibnamefont {Bark}}, \bibinfo {author}
  {\bibfnamefont {S.}~\bibnamefont {Ryu}}, \bibinfo {author} {\bibfnamefont
  {C.-B.}\ \bibnamefont {Eom}}, \ and\ \bibinfo {author} {\bibfnamefont
  {J.}~\bibnamefont {Levy}},\ }\href@noop {} {\bibinfo  {journal}
  {arXiv:1307.5557}\ }\BibitemShut {NoStop}%
\bibitem [{\citenamefont {Woodward}(1997)}]{Woodward:acsb97b}%
  \BibitemOpen
\bibfield  {journal} {  }\bibfield  {author} {\bibinfo {author} {\bibfnamefont
  {P.~M.}\ \bibnamefont {Woodward}},\ }\href {\doibase
  10.1107/S0108768196012050} {\bibfield  {journal} {\bibinfo  {journal} {Acta
  Crystallogr. Sect. B: Struct. Sci.}\ }\textbf {\bibinfo {volume} {53}},\
  \bibinfo {pages} {44} (\bibinfo {year} {1997})}\BibitemShut {NoStop}%
\bibitem [{\citenamefont {Linz}\ and\ \citenamefont
  {Herrington}(1958)}]{Linz:jcp58}%
  \BibitemOpen
  \bibfield  {author} {\bibinfo {author} {\bibfnamefont {A.}~\bibnamefont
  {Linz}}\ and\ \bibinfo {author} {\bibfnamefont {K.}~\bibnamefont
  {Herrington}},\ }\href {\doibase 10.1063/1.1744278} {\bibfield  {journal}
  {\bibinfo  {journal} {J. Chem. Phys.}\ }\textbf {\bibinfo {volume} {28}},\
  \bibinfo {pages} {824} (\bibinfo {year} {1958})}\BibitemShut {NoStop}%
\end{thebibliography}

%

\end{document}